# Mechanism of pressure induced amorphization of $SnI_4$: a combined X-ray diffraction – X-ray absorption spectroscopy study

Emiliano Fonda [1,*], Alain Polian [1,2], Toru Shinmei [3], Tetsuo Irifune [3,4], Jean-Paul Itié [1]

[1] Synchrotron SOLEIL, L'Orme des Merisiers, St Aubin BP48, 91192 Gif sur Yvette Cedex
[2] Institut de Minéralogie, de Physique des Matériaux et de Cosmochimie - CNRS UMR 7590, Sorbonne Université, 4 Place Jussieu, 75005 Paris
[3] Geodynamics Research Center, Ehime University, 2–5 Bunkyo-cho, Matsuyama 790-8577, Japan
[4] Earth-Life Science Institute, Tokyo Institute of Technology, Tokyo 152-8500, Japan

*corresponding author: fonda@synchrotron-soleil.fr

## ABSTRACT

We have studied the amorphization process of $SnI_4$ up to 26.8GPa with unprecedented experimental details by combining Sn and I K edge X-ray absorption spectroscopy and powder X-ray diffraction. Standards and reverse Monte Carlo extended X-ray absorption fine structure (EXAFS) refinements confirm that the $SnI_4$ tetrahedron is a fundamental structural unit that is preserved through the crystalline phase-I to crystalline phase-II transition about 7 to 10GPa and then in the amorphous phase that appears above 20GPa. Up to now unexploited Iodine EXAFS reveals to be extremely informative and confirms the formation of iodine iodine short bonds close to 2.85Å in the amorphous phase at 26.8 GPa. A coordination number increase of Sn in the crystalline phase-II appears to be excluded, while the deformation of the tetrahedral units proceeds through a flattening that keeps the average I-Sn-I angle close to 109.5°. Moreover, we put in evidence the impact of pressure on the Sn near edge structure under competing geometrical and electronic effects.

## INTRODUCTION

Pressure induced amorphization (PIA) is a quite general and fundamental phenomenon that has been observed in several one-, two- as well as tri-dimensional materials. There are various processes that may explain the origin of the order-disorder transition. In quartz at ambient temperature, the amorphization is due to a coordination change from 4 to 6 around the Silicon atoms, which leads to a new high pressure disordered polymorph[1-3]. In nano-anatase[4] ($nTiO_2$) the coordination increases from 6 to 7 like in the case of bulk anatase, where the high pressure form adopts the baddeleyite[5] structure (coordination 7). In this case the amorphous phase can be seen as a disordered form of the high pressure crystalline phase. In the case of semiconductors like Silicon[6] it has been proposed that amorphization corresponds to the crossover of a virtual melting line in the stability range of the solid phase. For GaP[7] or GaAs[8], an amorphous phase can be obtained by decreasing the pressure after a transition to the high pressure phase (Cmcm). The origin of the amorphous part of the sample is the existence of

"wrong" bonding (Ga-Ga bonds created in the high pressure phase) in the recovered phase. For the two first examples the amorphization is reversible, while for aromatic molecular crystals like benzene[9] or thiophene[10], an irreversible amorphization is observed under high pressure due to the break of the itinerant bond and the creation of a highly reticulated polymer.

Another family of molecular crystals exhibits reversible pressure induced amorphization: $AX_4$ compounds where A= Sn or Ge and X = Cl, Br, I [11, 12]. The amorphization has been already observed for $GeI_4$ and $SnI_4$, which is accompanied by a metallization. The mechanism proposed to explain this phase transformation is a charge transfer from intramolecular (A-X) to intermolecular bonds (X-X). At ambient pressure these compounds crystallize in a cubic structure where $AX_4$ molecules interact through Van der Waals forces (Fig.1).

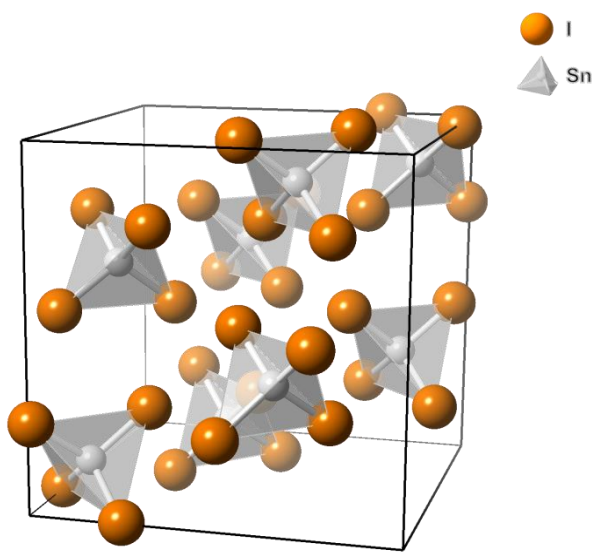

Figure 1. The structure of $SnI_4$ at ambient pressure.

At high pressure, charge transfer leads to the creation of conducting A-X-X-A-X…. chains (Fig.2). Such mechanism can effectively explain the amorphization and the metallization, but it has not been fully demonstrated due to the lack of crystallographic information.

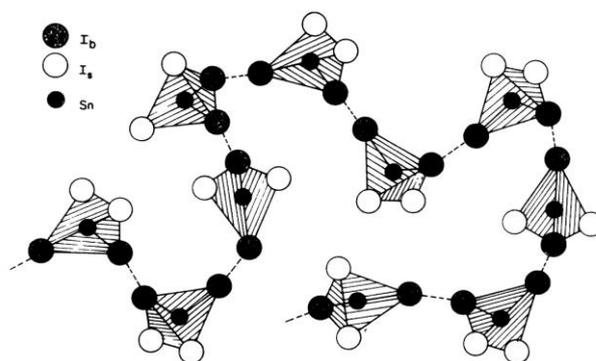

Figure 2 Proposed structural model for the high pressure amorphous phase of $SnI_4$, image reproduced from ref. [12]

For $GeI_4$ it has been shown by X-ray absorption spectroscopy (XAS) that the Ge-I distance increases when amorphization occurs[13]. This is in agreement with the proposed mechanism but an increase in the Ge coordination would lead to the same result. In fact, only XAS experiments under pressure at the halogen K edge could discriminate between charge transfer and coordination change.

$SnI_4$ has been extensively studied under high pressure first by optical absorption[14] and electrical conductivity[15], showing a closing of the gap and a metallic conductivity around 18.5 GPa. The subsequent X-ray diffraction [16, 17] or Raman scattering[18] studies have shown a progressive pressure induced amorphization accompanying the metallization. The amorphous phase is presented as formed by dimerized $SnI_4$ tetrahedra. A subsequent Mössbauer study[12], proposed that the amorphous phase above 20 GPa is composed of chains of $SnI_4$ tetrahedra bonded by iodine – iodine bridges, these chains explaining the metallic conduction of the compound. A first EXAFS (extended X-ray absorption fine structure) study[19] has been performed using sintered boron carbide anvils. The result did not agree with the proposed "chain model" deduced from Mössbauer experiment, but proposed that the amorphous phase is formed by distorted $SnI_4$ molecules rotated in such a way that tetrahedral units are formed among the iodine atoms, with tin atoms as interstitials. The next studies using X-ray diffraction proposed a new scheme. First, a study up to 153 GPa[20] observed an intermediate structure, called CP-II between the ambient pressure phase (CP-I) and the amorphous one, with a transition beginning around 7.2 GPa. The amorphization begins around 15 GPa, and a new crystalline phase, cubic, CP-III, appears at 61 GPa, and remains unchanged up to 153 GPa. In this new high pressure crystalline phase, the first neighbors distance is about 3 Å, which tends to prove the dissociation of the $SnI_4$ molecules. During decompression, the amorphous phase is recovered at 30 GPa, transit to another amorphous structure (Am-II) at 1.8 GPa and finally to the CP-I at 0.4 GPa. Nevertheless, the authors did not infirm the dimer model for the amorphous structure. A dependence on the compression history is recognized in another study [21]: a second compression moves the frontier between Am-II and Am-I phases from 3 to 7 GPa upon compression, while in first compression CP-I was followed by CP-II and then Am-I. A following paper [22] studied specifically the structure of the amorphous state Am-I by X-ray scattering and it concluded that in this form of $SnI_4$, the molecules dissociate leading to a state with great similarities with metallic glasses such as Ni or Fe, because the position of the maximum of radial distribution function does not coincide with the intramolecular atomic distances. The decomposition is supposed to occur above 25 GPa, and the coordination number deduced between 35 and 55 GPa, considering that Sn and I atoms are indistinguishable, is about 10. Finally, very recently a study has been devoted to the mechanism of the amorphization [23] using *ab initio* molecular dynamics calculations and nuclear resonant inelastic X-ray scattering (NRIXS). The conclusion is that the PIA is due to a mechanical instability (the shear elastic modulus $C_{11} - C_{12}$ -2P becoming negative at 10.4 GPa), and is not due to the metallization, which is correlated with the CP-I – CP-II phase transition instead. The structures of the CP-II and of the amorphous phases are also examined in this work. It is proposed that in Am-I individual penta-atomic tetrahedra do not exist anymore: tin atoms are six-fold coordinated to four iodine atoms at "short" distances (2.93 Å), and a pair of iodine atoms at 3.3 Å. It is proposed that the CP-II phase belongs to the $P\bar{1}$ space group and is composed of edge sharing Sn-I octahedrons disposed in a zigzag ribbon along the b-axis. Very recently, an EXAFS study has been published as a function of pressure and temperature up to 9 GPa and 900 K [24], in the goal of determining a possible deformation of the $SnI_4$ tetrahedron in the solid phase before the melting, where a transformation from the $T_d$ to the $C_{3v}$ point group has been observed [25]. From these results, the coordination number does not change up to 9 GPa, and the intramolecular distances are in good agreement with the published results [26], giving a "molecular modulus" of 157 GPa at ambient pressure and temperature. On the contrary no result concerning the amorphization process is presented.

There is therefore for the moment a quite blurry landscape of the amorphization process and amorphous state from the local point of view. X-ray absorption spectroscopy (XAS) is a technique very well adapted for such a study, since it probes the local environment of a chosen atomic specie, giving the coordination number of the tested atom, the neighbors distances and the disorder degree (Debye-Waller factor). In this manuscript, we apply classic [27, 28] and recent advances in reverse Monte Carlo (RMC) EXAFS simulations [29-31] to the room temperature pressure induced amorphization of $SnI_4$. RMC-EXAFS simulations allow us to describe with unprecedented details the complexity of the local structure modifications passing from the low pressure CP-I to the high pressure amorphous phase.

We have explored the behavior of $SnI_4$ as a function of pressure by combining X-ray diffraction and X-ray absorption spectroscopy at the Sn K-edge (29.2 keV) and at the I K-edge (33.169 keV) on the same setup, under exactly the same conditions. These experiments enable the exploration of the neighborhood of both atomic species and hence to characterize the approach and the formation of the amorphous phase.

EXPERIMENTAL

Until recently XAS experiments under pressure at edges above 11 keV were technically extremely challenging in a DAC. There were two possibilities: the first one was to use opaque cubic $B_4C$ as anvils [32], which prevents the use of the ruby pressure scale [33]. The second one was to make the measurements at two different angles, in order to move the diamond Bragg peaks to other positions, and hence to allow the determination of the EXAFS spectrum by combination-subtraction of the two spectra[34]. In fact, this technique was only applicable for energies where the number of Bragg peaks is not too large, *i.e.* typically below 12 keV. This is especially true at the energy of Sn and I K edges (29 - 33 keV), energies at which an EXAFS scan (about 1000 eV) can encompass a very large number of diffraction peaks.

Since 2003, nanopolycrystalline diamonds (NPD) with no binder materials have been synthetized from graphite at high pressure and high temperature [35, 36] and they are polished to produce anvils. These NPD consist of nanocrystals (typical size some tens of nanometers), randomly oriented. Hence, their diffraction pattern is close to that of an ideal powder contributing only as a smooth background to the absorption spectrum.

EXAFS spectra have been measured at the SAMBA beamline[37] of the SOLEIL synchrotron (Gif sur Yvette, France) in transmission mode with two Ar filled ionization chambers. Diffraction patterns have been measured at 29 keV (200 eV below Sn K edge) with a Rayonix MARCCD large area detector placed after the sample. The second ionization chamber ($I_1$) and the MARCCD were alternatively placed after the sample at each pressure step. A membrane diamond anvil cell (DAC) equipped with 400 μm culet diameter nanopolycrystalline anvils has been used. The experimental volume was a 200 μm diameter hole drilled in a 150 μm thick Re gasket pre-indented to 25 μm. The usual SAMBA beam footprint is 300x200 μm² (horizontal times vertical dimensions), thus beam has been defocused and a pair of JJ-X-rays slits has been positioned close to the sample to reduce the beam size below the size of the gasket hole. Between the slits and the sample, a short ionization chamber has been placed (IC-Plus 50 by FMB-Oxford) to measure the incoming flux ($I_0$), while the monochromator stabilization control (I200, FMB Oxford) was performed on another ionization chamber placed before the slits. The monochromator was equipped with a pair of Si (220) crystals. Harmonic rejection was performed by using a pair of Pd coated (50 nm) Si mirrors placed before and after the monochromator at a grazing angle of 1.5 mrad. Diffraction data calibration has been performed measuring a $CeO_2$ powder.

Radial integration of diffractograms and usual corrections have been performed with Fit2D [38, 39]. Non-linear least square fits of EXAFS spectra have been performed with the Demeter package [27] on the basis of Feff6 [40] theoretical standards on the k range 3.5 – 14 Å$^{-1}$ except for 22.6 and 26.8 GPa where the range has been shortened and moved closer to the edge due to the larger disorder: 3.0 – 10.0 Å$^{-1}$. Evolutionary RMC simulations of EXAFS data were performed with the EvAX code [29, 30, 41] employing Feff85L[42] theoretical standards; calculations were performed thanks to the Synchrotron SOLEIL high performance computing (HPC) facility. EvAX calculations were performed on 3x3x3 cells with a fixed lattice parameter $a$ deduced from diffraction data. $S_0^2$ and $E_0$ values were fixed during refinements and have been deduced from standards and very limited stepwise optimization respectively.

## RESULTS AND DISCUSSION

Diffraction data evidence a progressive amorphization as indicated by the broadening and loss of intensity of the $SnI_4$ peaks (open circles in Fig.3) above 12.9 GPa; we recorded diffractograms during decompression too and they witnessed the reversibility of the process with a distinct hysteretic behavior (decompression at 6.5 GPa remains in the amorphous phase while compression up to 6.5 GPa remains in the crystalline phase). Diffraction provides a direct evidence for lattice compression as evidenced by the shifts of all $SnI_4$ peaks at larger angles; obtained lattice parameters (*a*) are reported in Table 1; values have been extrapolated above 15.4 GPa by fitting a Murnaghan equation of state [43], which for a cubic crystal may be written as $\frac{a}{a_0} = \left[1 + \frac{B_0'(p-p_0)}{B}\right]^{-\frac{1}{3B_0'}}$, where a and $a_0$ are the cubic lattice parameter at pressure p and $p_0$ and $B_0$ and $B_0'$ are the bulk modulus and its pressure derivative at pressure $p_0$.

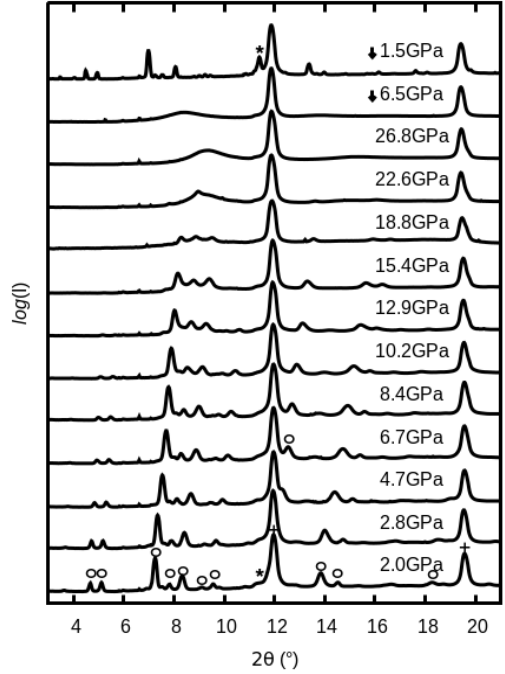

Figure 3 X-ray diffractogram obtained by sector integration as a function of applied pressure. SnI$_4$ pristine structure peaks are indicated by a (o), Re gasket (*) and diamond (+). Down pointing arrows have been placed close to the pressure values obtained after decompression. All curves are vertically translated for clarity.

Tableau 1 Lattice parameter measured or extrapolated(#) and first shell Sn-EXAFS fitting results. N is the coordination number, R the interatomic distance, $\sigma^2$ is the EXAFS Debye Waller factor.

| P (GPa) | a(Å) | | N | R (Å) | $\sigma^2$ (10$^{-3}$ Å$^2$) |
|---|---|---|---|---|---|
| 2.0 | 11.79(1) | Sn-I | 4.0(2) | 2.668(2) | 3.2(3) |
| 2.8 | 11.65(1) | Sn-I | 4.1(2) | 2.664(2) | 3.1(2) |
| 4.7 | 11.34(1) | Sn-I | 4.4(3) | 2.653(3) | 3.4(3) |
| 6.7 | 11.11(1) | Sn-I | 4.3(3) | 2.644(3) | 4.0(3) |
| 8.4 | 10.98(1) | Sn-I | 3.7(2) | 2.639(2) | 3.1(3) |
| 10.2 | 10.81(1) | Sn-I | 3.7(2) | 2.632(2) | 3.7(3) |
| 12.9 | 10.64(1) | Sn-I | 3.0(4) | 2.649(8) | 4.0(9) |
| | | Sn-I | 1.1(4) | 2.84(2) | |
| 15.4 | 10.46(1) | Sn-I | 3.5(8) | 2.71(2) | 7(2) |
| | | Sn-I | 1.8(4) | 2.86(3) | |
| 18.8 | 10.3(1) # | Sn-I | 3.2(1) | 2.70(1) | 9(3) |
| | | Sn-I | 1.3(9) | 2.82(6) | |
| 22.6 | 10.2(1) # | Sn-I | *9(3)* | 2.92(3) | 37(3) |
| 26.8 | 10.0(1) # | Sn-I | *8(1)* | 2.90(1) | 23(3) |

Measured and extrapolated *a* values have been used in the following to build starting models for EXAFS fittings. Data are displayed in Fig. 4. It must be noted that these EXAFS spectra are of exceptional quality considering the large k and pressure range and the fact that they have been measured at high energy in a diamond anvil cell: very few spurious peaks punctuate curves thanks to the use of nanopolycrystalline diamonds. For the same reason we were able to use the same cell alignment for Sn and I K edges (29.2 and 33.2 keV).

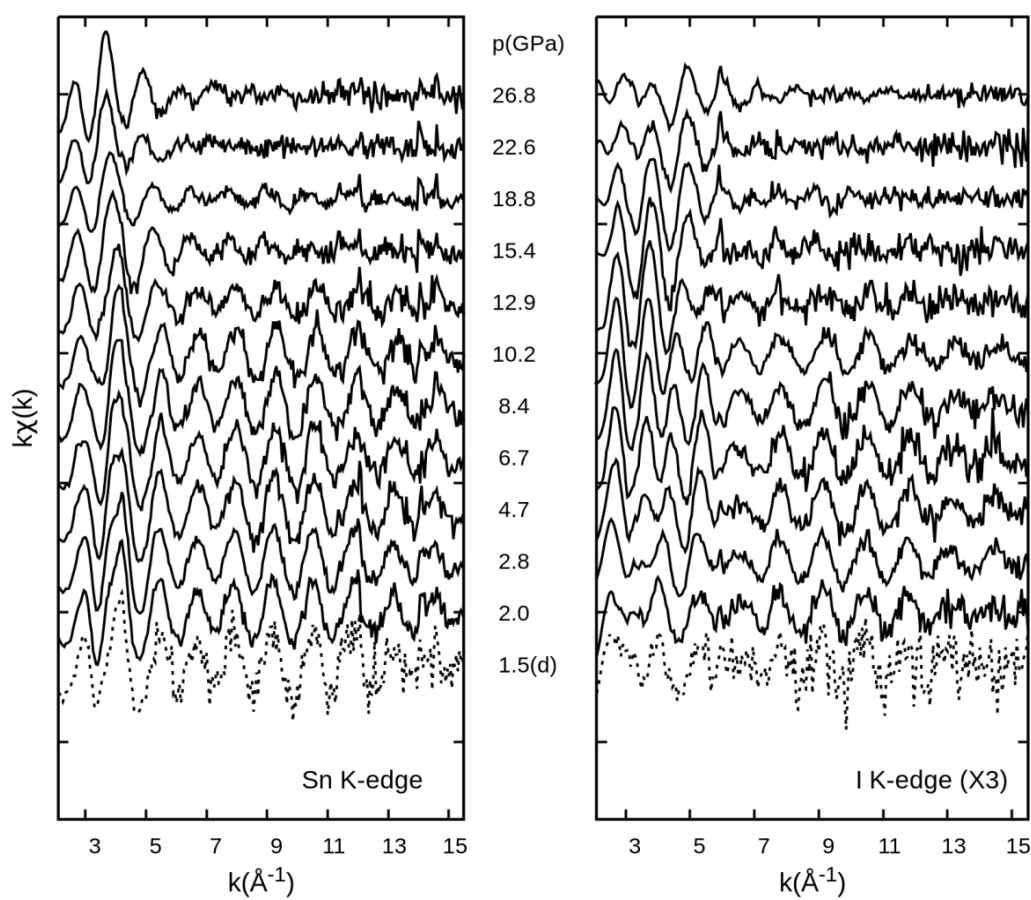

**Figure 4** Raw EXAFS data obtained at the K edges of Sn and I as a function of pressure. A wider data range has been measured, shown is the exploited data range.

It is much easier to understand the effect of pressure on the local structure by inspection of the EXAFS Fourier transforms (FT, Fig.5): Sn is directly coordinated to 4 iodine and each iodine is initially only coordinated to one tin, thus each FT has only one notable peak at the apparent distance of the Sn-I bond minus the atomic phase shift [44]. Increasing pressure, Sn surroundings change little below 12.9 GPa, while around I a second peak appears at larger distances, grows and moves to lower distances. This new peak can only be due to the approach of iodine belonging to another SnI$_4$ unit. Above 12.9 GPa the Sn–I bond apparently elongates (the first peak in both panels of Fig. 5 shifts right) and progressively broadens up to 26.8 GPa where it appears to sharpen again. The process appears progressive and complex. As from diffraction data, there is not a definite amorphization pressure.

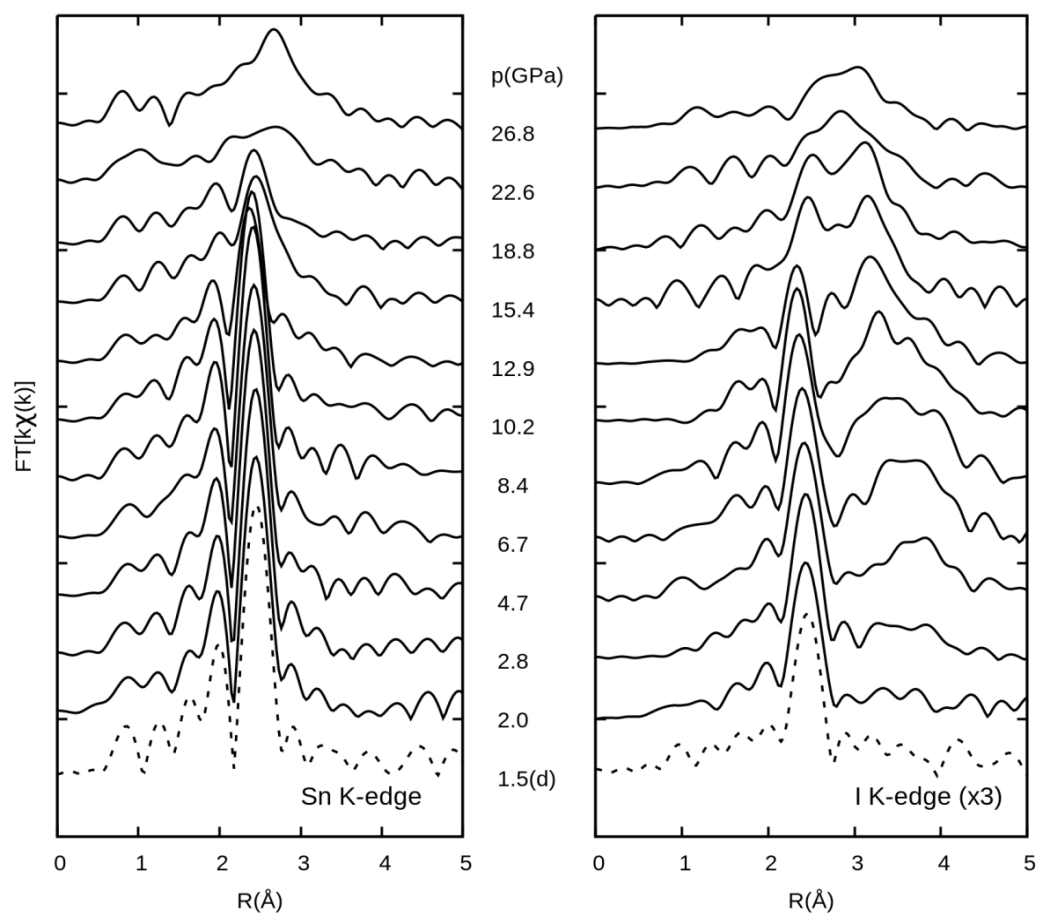

**Figure 5** Fourier transforms (k-weighted, modulus) of K-edge EXAFS spectra of Sn and I as a function of pressure. The (d) symbol indicates the spectra obtained after decompression. All curves have been shifted upwards for clarity.

Numerically fitting the Sn K edge EXAFS provides reliable results for Sn-I distances and coordination numbers up to 18.8 GPa (Table 1.): the total Sn coordination number is always close to 4 as expected, but starting from 12.9 GPa, two subshells can be distinguished. Taking the shorter distance, it shortens up to 10.2 GPa and then clearly elongates as the pressure is increased together with an increase of the pseudo Debye Waller factor ($\sigma^2$). Above 18.8 GPa the overall distance reaches a plateau, $\sigma^2$ passing through a maximum at 22.6 GPa and the coordination number seems to increase up to 8. Actually, these numbers must be taken with great caution since the fitting model is not adequate for such a large disorder [45], particularly if the pair distribution function has more than one peak, case where even extending the fitting to the third and fourth cumulant does not help. Fitting the I K edge EXAFS is even more complex since it may require too many parameters due to overlapping contributions. A different approach is thus required as provided by RMC fitting: we used the cell parameters obtained by XRD (extrapolated and measured) to build 3x3x3 super cells based on the low pressure pristine structure and then fit EXAFS data of both I and Sn edges simultaneously with the EvAX code [29, 30, 41]: the structure is refined by moving atoms with a reverse Monte Carlo method improved by the introduction of an evolutionary algorithm under periodic boundary conditions. Data and theory wavelets transforms [31] are compared and the best match is sought through the evolutionary algorithm described by Timoshenko *et al.* [29]. The main advantage consists in correlating a real structure (the starting point being well known) with EXAFS data and then being able to statistically analyze the obtained super cells. These have no meaning when locally inspected, but the average quantities make sense since the EXAFS spectra are the average over all Sn and I sites in the super cells (3x3x3 cell containing 1080 atoms), at the same time we can try to obtain additional information, distinct from what is obtained by fitting the analytical EXAFS formula: we can distinguish between intra and intermolecular I–I distances, shown in Fig.6, and obtain the distribution of the I–Sn–I angle. Both quantities being related to I–I and Sn–I distances as well as multiple scattering paths and constrained by the cell size and periodic boundaries conditions.

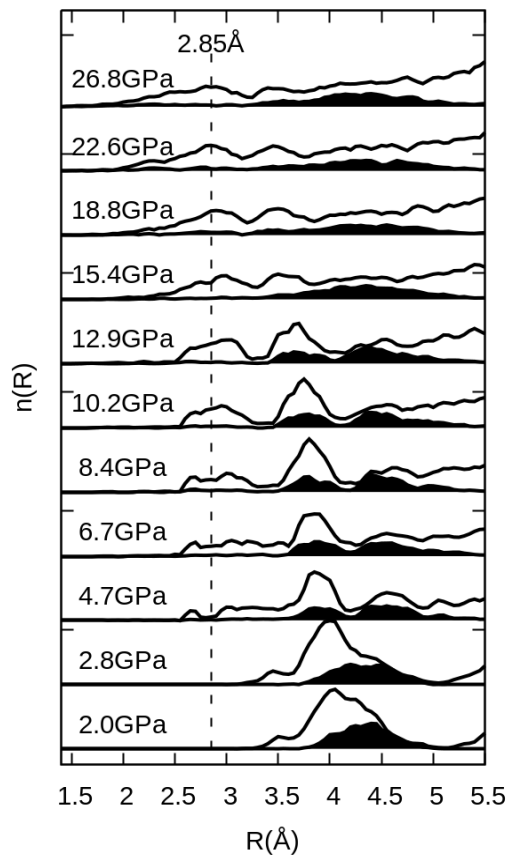

Figure 6 The distribution of I-I pairs distances divided between intramolecular (black filled curve) and intermolecular pairs (continuous line) as obtained by the analysis of the RMC EXAFS fits.

The filled curved in Fig. 6 describes well the transition from a tetrahedral coordination with identical I–I intramolecular distances around one Sn center and the splitting in two sub shells starting at 4.7 GPa together with the onset of an intermolecular I–I contribution (continuous black line, Fig. 6) below 3Å that progressively shifts at smaller distances. At 15.4 GPa, the intramolecular I–I distribution flattens while the shortest intermolecular distance is still well defined. This picture bring us back to the model proposed by Pasternak and Taylor in 1988 [12] and illustrated in Fig. 2, tetrahedral units form chains through a charge transfer from Sn to I. Sn–I distance elongates while the I–I inter-tetrahedra bond forms. This precedes amorphization and causes the resistivity drop and the transition from an insulating to a metallic phase [11]. Iodine EXAFS is crucial to obtain such details as well as the combining of both Sn and I information to properly constrain results for the first time. I–Sn–I intramolecular angular distribution comforts our conclusions (Fig.7).

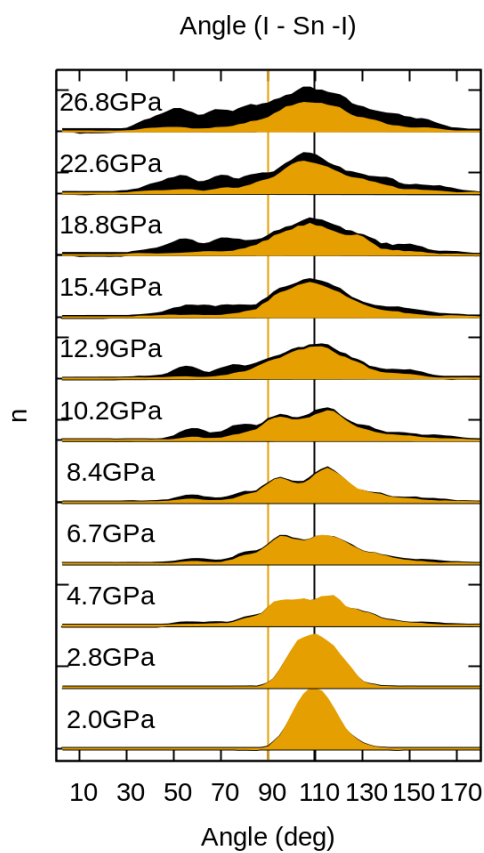

Figure 7 Distribution of the I-Sn-I angle as obtained by the analysis of the RMC EXAFS fits for Sn-I pairs up t 3.5Å: the intra molecular angles distribution is represented by the yellow shaded surface where the black one counts for all I-Sn-I angles. The low angular tailing is clearly due to intermolecular angles at these pressures.

The average angle is at about 109.5° at 2 GPa as it should be for a tetrahedron, while it progressively splits from 4.7 GPa up to 12.9 GPa; then the distribution progressively flattens. Nonetheless the distribution remains centered around 109.5° indicating that tetrahedral units are still the majority even in the amorphous phase. As already done for I–I pairs we can distinguish intra molecular and intermolecular I–Sn–I angles as shown in figure 7: this well confirms the existence of two distinct contributions up to amorphization, the intra molecular angles remaining close to a tetrahedral, even if strained and deformed, geometry. The intermolecular contribute mostly at smaller angles, meaning that they approach the Sn core in between the coordinated iodines. At the highest pressures as amorphization proceeds, the frontier between intra and intermolecular blurs, but the process is progressive on the local scale. Unfortunately, only measurements at higher pressures could confirm what kind of coordination change is taking place before CP-3 appears above 60GPa.

Liu et al. [23] reported on the base of XRD data and Molecular Dynamics, that the CP-2 phase should appear between 10 and 12 GPa. In the CP-2 they proposed an expansion of the Sn coordination to 6 with a distorted tetrahedral coordination and the coordination of two more I atoms at about 3.3Å (Sn–I) plus the first four at about 2.93 Å towards 20 GPa. We well confirm the deformation and our findings agree extremely well with the coordination of four I at 2.93 Å as can be deduced from Table 1 and the Sn – I pair distribution drawn in Fig. 8; however, we have shown an intra average I–Sn–I always close to 109.5° and there is no trace of an expansion of the Sn coordination to 6, and definitely not a relevant contribution of I around 3.3 Å (Fig. 8).

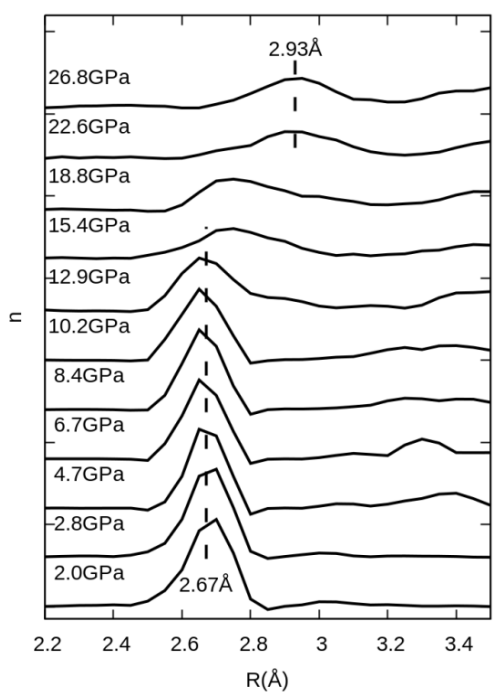

**Figure 8** Distribution of the Sn-I pairs as obtained by the analysis of the RMC EXAFS fits. Vertical dashed lines are guides for the eye at the two labeled distances of 2.67 and 2.93Å (see text).

X-ray absorption near edge structure (XANES) is the part of the absorption spectrum across the photo ionization threshold and it is represented in Fig. 9. At the K edge of Sn it is characterized by a single peak originating from *1s → 5p* transition, the dipole selection rules imposing the *p* character of the final state. The *4d* shell being formally full even for the highest Sn oxidation states, in $SnI_4$ only a partial charge transfer is expected between Sn and I and involving *5sp* orbitals. Indeed this was already proposed to explain the Mössbauer signal observed by Pasternak et al. in 1988 [12] for $SnI_4$ under high pressure: a second type of $^{119}$Sn, named $Sn_h$, where *h* stands for high pressure, appears at about 10 GPa and is present as 100% at 25 GPa.

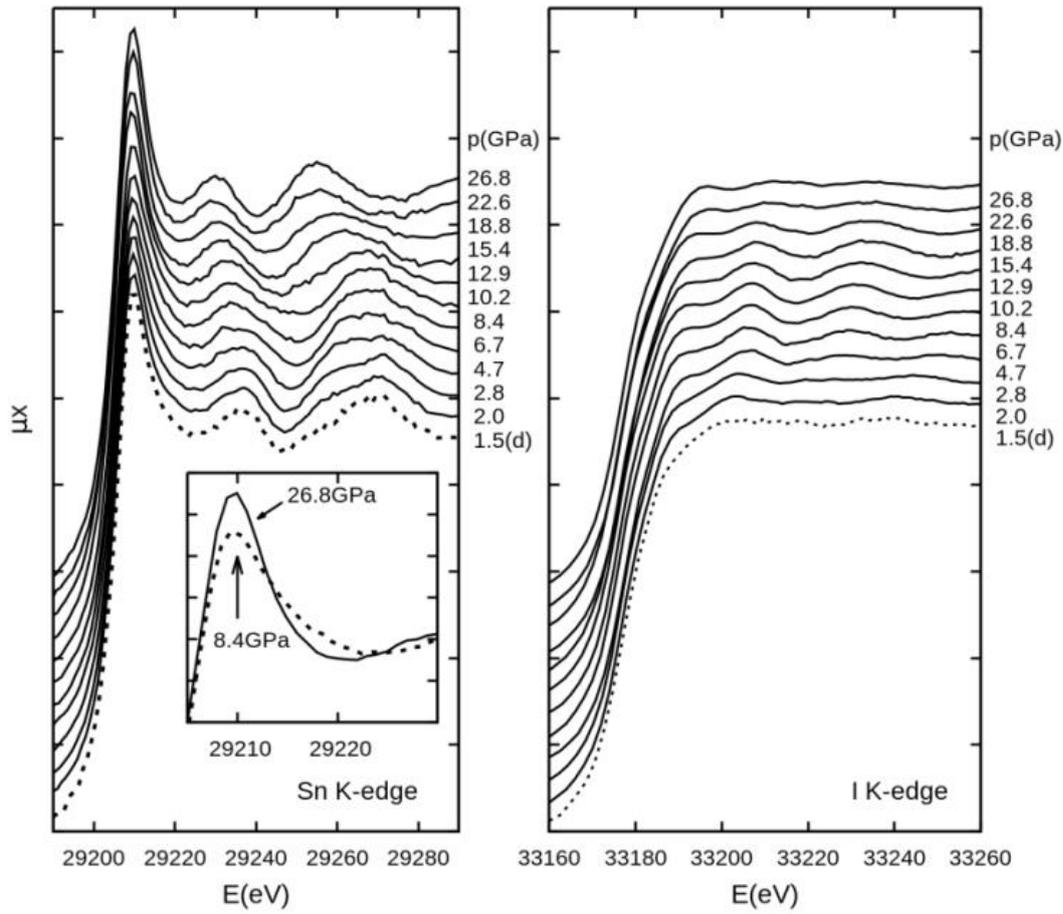

Figure 9 Near edge (XANES) region of K edge XAS spectra of Sn (left panel) and I (right panel) as a function of pressure. The (d) symbol indicates the spectra obtained after decompression. The inset in the left panel shows the white line intensity change between the extremes at 8.4GPa (minimum) and 26.8GPa (maximum intensity), while no edge shift can be evidenced larger than experimental error. Except for the two in the inset, all other curves have been shifted up for clarity.

A similar trend was found for the $^{129}$I Mössbauer signal presenting a second population at high pressure reaching from 10 GPa to 17 GPa a 55% population as a saturation value. In Fig. 10 we report the white line intensity as a unitless number. It represents the top of the normalized XANES peak and we compare it with the shortest Sn–I distance: the intensity lowers with increasing pressure reaching a minimum at 8.4 GPa and then increases again and maximizes at 26.8 GPa.

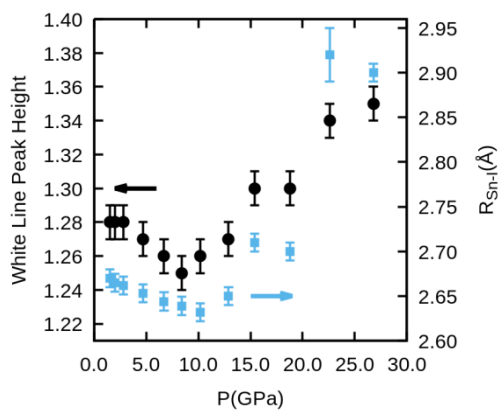

Figure 10 Filled circles represent (left axis) the Sn K edge XANES top of the white line versus pressure; the light squares (right axis) report the Sn-I shorter distance (when two were found in fit). The two curves exhibit very similar trend, but white line intensity curve has a minimum at 8.4GPa right before that of the Sn-I distance that is minimum at 10.2GPa.

It is evident that the minimum of the XANES intensity precedes that of the Sn—I bond and they cannot be simply directly related. K edge XANES is an image of the density of empty states of type $p$ and depends on filling and symmetry. Symmetry of the $SnI_4$ units can be, at least partially, gauged by the I–Sn–I angle (Fig.7) that splits in two departing contributions up to 8.4 GPa and then converge again at higher pressures. We can propose a qualitative interpretation based on two competing effects: first, the deformation of the $SnI_4$ units from $T_d$ geometry lowers the XANES peak intensity below 8.4 GPa, while I···I bonds formation associated to charge transfer and Sn—I bond elongation increases it above 8.4 GPa. Iodine XANES is much less informative (the $p$ orbitals being almost filled) and is not discussed, either we cannot significantly compare the edge shift of Sn or I within this experiment since they are below the experimental error.

## CONCLUDING REMARKS

The transition from the low pressure crystalline state to the amorphous high pressure state has been analyzed under a new point of view and with unprecedented experimental detail thanks to the combination of XRD, and the X-ray absorption spectra of all elements in the cell. A similar approach had been already extremely valuable in the past to describe a complex system like Ru doped $Mn_2Ga$ [46] and confirmed here its importance.

During the amorphization the structural unit never departs from tetrahedron, $SnI_4$ units approach with almost no shrinking, then units distort and shrink from 4.7 to 8.4 GPa and finally a number of I···I bonds forms at 10.2 GPa causing the before compressed Sn—I distance to elongate. Amorphization takes place afterwards starting from 15.4 GPa and being complete at 22.6 GPa. In the amorphous phase Sn—I and I—I shorter distances converge, being about 2.93 and 2.85 Å respectively, shortest I—I being even shorter than Sn—I.

The effect of structural changes in SnI$_4$ on Sn K-edge XANES is reported here for the first time, most likely thanks to the good data quality obtained by the combination of nanocrystalline diamonds and the use of SAMBA.

### ACKNOWLEDGMENTS

Authors thank Synchrotron SOLEIL for providing beam time and G. Alizon for his highly appreciated technical support. E.F. thanks P. Martinez for introducing him to the Synchrotron SOLEIL HPC facility and Synchrotron SOLEIL for allowing computation time. EF warmly thanks J. Timoshenko and the EXAFS Spectroscopy Laboratory of the Institute of Solid State Physics (Riga, Latvia) for providing and maintaining the code EvAX.